# Complex fluids in the animal kingdom

Patrick A. Rühs[1], Jotam Bergfreund[2], Pascal Bertsch[2], Stefan Gstöhl[2], Peter Fischer[2*]

[1]Department of Bioengineering, University of California, Berkeley, CA, USA

[2]Department of Health, Science, and Technology, Institute of Food Nutrition and Health, ETH Zurich, Switzerland

*Corresponding author: peter.fischer@hest.ethz.ch

**Abstract**

Animals have evolved distinctive survival strategies in response to constant selective pressure. In this review, we highlight how animals exploit complex flow phenomena by manipulating their habitat or by producing complex fluids. In particular, we outline different strategies evolved for movement, defense from predators, attacking of prey, and reproduction and breeding. From the slimy defense of the notorious hagfish to the circus-like mating spectacle of leopard slugs, we unveil remarkable correlations within the flow behavior and biological purpose of biological complex fluids. We discuss recurring phenomena, propose flow behavior for undescribed complex fluids, and put these in context with the animal's survival strategy. With this review, we hope to underline the importance of complex fluids and material flow in the animal kingdom.

# Animal Rheology

**Introduction**
Animals are under constant selective pressure to adapt themselves to their surrounding environment. Countless survival strategies have evolved and enabled animals to survive, colonize, and reproduce successfully in the most diverse habitats on earth[1]. From penguin species climate-adapted size differences[2] to the food-adapted Darwin finches[3], an overwhelming array of examples suggests that selective pressure continues to drive natural selection and, if successful, results in well-adapted animals that integrate into an ever-changing environment[4]. Selective pressure derives from the entirety of environmental factors acting on animals that can be biotic or abiotic. Biotic factors include all organisms in an individual's environment, the individual itself, and the resulting consequences, such as competition for food supply or physical and chemical changes in the environment. Abiotic factors are of chemical or physical nature; they are non-living factors such as temperature, humidity, light, nutrients, and the mechanical properties of their surrounding materials. The animals' close relationship with the surrounding materials strongly influences their behavior, e.g. locomotion, while also providing them with food and shelter[4]. These intriguing material-animal relationships featuring optimized mechanisms, unprecedented material properties[5,6], and matching biological behaviors are bio-inspiring a large community[7]. We want to demonstrate how the understanding of soft material – animal interaction offers a new quantitative perspective to animal behaviour in the context of evolution.

In evolutionary biology, materials are distinguished as of endogen or exogen origin. Endogen materials originate from the animal itself, whereas exogen materials originate in the environment. In turn, both can fulfill animal body functions and exploitation. Endogen abiotic material originates from an animal (endogen), however, it is not formed by a biological process (abiotic), e.g. a humpback whales blowing a "net" of air bubbles to trap fish[8]. In contrast, exogen biotic material is formed by living organisms but not by the individual that utilizes it. Insects, for example, feed on and inhabit cow dung, as it provides shelter and nutrients for their larvae[9]. Endogen biotic material is formed by a biological process and originates from the organism that utilizes it, such as saliva for lubrication and digestion. Thus, the materials often link directly or indirectly to an adaptive behavioral trait, relieving a species from selective pressure.

Such materials, regardless of their living or non-living state, can be classified by their mechanical properties into hard and soft materials[10]. Hard (i.e., solid) materials of exogen origin generally do not change over time or as a function of applied stress and offer only little physical response to be exploited by animals. In contrast, solidified endogen material can serve the body function and be used by the animal for a specific behavior. Darwin finches for example secured their ecological niches by evolving various sizes and shapes of their keratin-based beaks to feed on different sources. The structure-property relationships of these beaks were optimized by nature on different length scales through genetic drift, natural selection, and adaptable behavioral traits, resulting in an evolutionary advantage. Another example is given by the stone-eating worm in the Philippines, *L. abatanica*[11] that has hard denticles that resemble a grinder. Such solidified biological materials are often strong and durable and can be described directly by their elastic or plastic properties.

In contrast, fluids, also called complex fluids when consisting of multiple phases [12], are easily affected by external forces and often have fascinating material properties as a function of time, stress, and temperature. Complex fluids are multiphase systems consisting of liquid, solid, and gaseous components. The simplest way of describing the mechanical properties of fluid is by its viscosity, a measure of the internal friction. For Newtonian fluids, the viscosity is constant regardless of the external acting forces. Complex fluids, as the name suggests, exhibit a more complex, non-Newtonian flow behavior. For example, the viscosity of a complex fluid may increase or decrease within seconds under mechanical stresses. Also, many complex fluids exhibit elastic (solid-like energy storage) and viscous (liquid-like energy dissipation) properties at the same time, and are thus defined as viscoelastic[13]. This broad spectrum of flow behaviour of soft materials offer a variety of mechanical properties to be exploited by animals.

To demonstrate the importance of complex fluids and flow phenomena in the animal kingdom, we highlight species that exploit complex fluids as part of their survival strategy. In detail, we review how animals cope with and utilize exogen or endogen, biotic or abiotic complex fluids as a competitive advantage for movement, prey, defense, and reproduction (Figure 1). We demonstrate how animals learned to manipulate their habitat, e.g., sandfish and crabs that exploit the granular rheology of sand (exogen abiotic) for their advantage and showcase exotic animals that produce unique bio-fluids (endogen biotic), from deep-sea hagfishes to subterranean velvet worms. For unknown and uncharted phenomena, we suggest the potential rheological properties of the involved complex fluids. This review aims to assess complex flow behaviors used by animals in nature, compare and reveal the effect of material properties on biological behavior, and demonstrate the importance of rheology in nature.

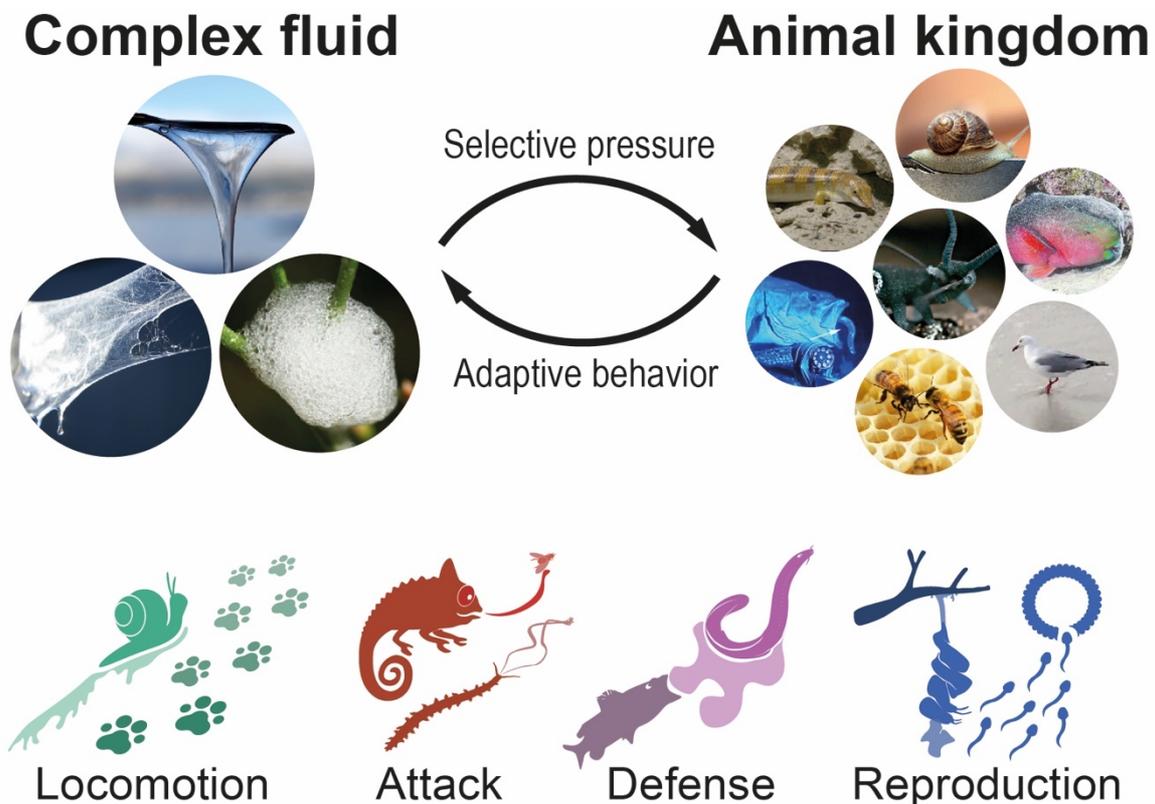

**Figure 1: Materials link to an adaptive behavioral trait to relieve a species from selective pressure.** Animals utilize or manipulate endogen or exogen, biotic or abiotic materials in fluid or solidified form for locomotion, attack, defense, and reproduction to improve Darwinian fitness.

**Rheology for locomotion**

Since the first animals set foot (or fin) on land 400 Mio years ago, animals colonize and adapt to all possible habitats. Different surrounding media (water, soil, air) wherein or whereon animals are moving has resulted in various specialized locomotion. For instance, in mammals, the same initial set of appendicular bones evolved into specialized limbs like fins, wings, hoofs, hands, and feet to enable different forms of locomotion[14]. With this 'mechanical' toolset, animals can act on their surroundings to propel themselves forward. Fluids, such as air or water, provide a strain-rate-independent viscosity (Newtonian). In other words, the stress to strain-rate ratio remains constant. Elsewhere, solidified soils respond linearly to stress or strain, resembling a linear spring mechanically (Hooke's law) until they yield. In contrast, complex fluids like sand or mucus exhibit strain and time dependent flow behavior. Herein, we demonstrate how gastropods secrete endogen viscoelastic locomotion aid and lizards exploit granular rheology to "swim" in sand (Figure 2).

*Movement by viscoelasticity.* Terrestrial gastropods, such as slugs and snails, crawl with a single foot by a mechanism called adhesive locomotion. This propulsion is powered by muscular waves that propagate along the ventral surface of the foot from tail to head[15,16]. These periodic contraction-relaxation waves (depicted in Figure 2a) are transmitted to the ground by a thin layer of viscoelastic mucus secreted by the animal[17]. The nonlinear rheology of the mucus enables the gastropods to propel without fully detaching from the solid ground by a stick-and-release mechanism. The mucus further provides adhesion to climb walls and crawl across ceilings, i.e. gastropod mucus requires specific rheological and adhesive properties[18,19].

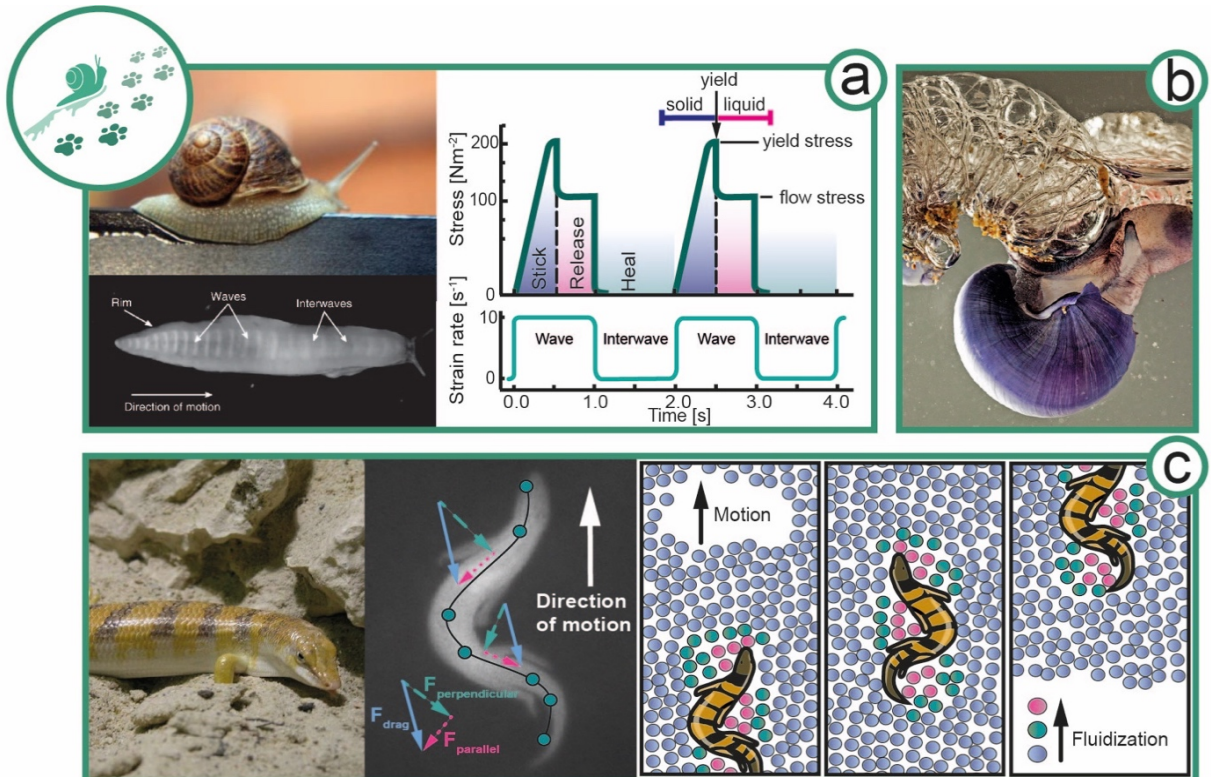

**Figure 2: Locomotion in the animal kingdom via complex fluids.** (a) Facilitated locomotion via mucus viscoelasticity by slugs. Plan view of the ventral surface of the foot of a slug moving over a glass surface from left to right[20]. Rheological properties of pedal mucus

of *Ariolimax Columbia* during locomotion (re-drawn from Denny *et al.*[17]). (b) The violet sea snail (*Janthina janthina*) uses complex fluids to form a foam raft for movement[21]. (c) A sandfish (*Scincus scincus*) on sand; an X-ray image of sandfish undulatory movement in glass beads; and a schematic illustration of the fluidization of sand particles surrounding the sandfish. The sandfish opposes strong transverse forces to his longitudinal axis to create a net forward thrust by relying on interparticle Coulomb friction. Rights and permission: (a) Snail image: UC3M Department of Thermal and Fluids Engineering, Carlos III of the University of Madrid. Bottom view of snail: adapted with permission from Lai *et al.*[20] (Copyright 2019 Company of Biologists Ltd.). (b) Picture of violet sea snail kindly provided by Dimitris Poursandinis. (c) By Wilfried Berns - Wilfried Berns / Tiermotive.de (CC BY-SA 2.0.) X-ray image of sandfish kindly provided by Daniel Goldman.

Gastropod mucus consists mainly of a mucin-like, protein-polysaccharide complex, similar to glycoproteins and glycosaminoglycans in vertebrates, which forms a physically crosslinked gel responsible for mucus viscoelasticity[22]. From a rheological point of view, mucus requires (i) a solid-like elasticity at low stresses, (ii) a high and sharp yield point with a transition to a highly viscous and shear-thinning liquid and (iii) a fast thixotropic recovery of network structure after stress release. Thus, a movement cycle, as visualized in Figure 2a, starts with elastic solid-like mucus. The approaching muscular wave shears the mucus, thereby increasing the stress. At the yield point, the mucus structure breaks and liquefies. During the interwave (no shear), the mucus network recovers its elastic, solid-like properties[17–19]. The movement cycle generates an asymmetric shear force under the foot with a net forward component that propels the animal. The stick-and-release enables terrestrial gastropods to move over and adhere to a wide variety of surfaces[20].

Adhesive locomotion is the most energy-intensive form of movement, approximately 12 times more costly than running. Surprisingly, the largest fraction of the energy is required for the production of mucus and not for muscle activity[23]. The energy input is well invested, as the rheological behavior beneficially enables snails to adhere to surfaces, for example, to climb up trees and overhangs. At the same time, mucus production and adhesion might limit the speed of snails. This well-defined balance between stick and slip demonstrates that the tailored rheology of mucus is optimized for gastropod propulsion[24].

*Passive movement by foam induced flotation.* A particularly low energy form of locomotion is passive flotation in aquatic environments. Violet sea snails (*Janthina janthina*) use complex fluids to form a foam raft for movement, as depicted in Figure 2b[21]. The sea snail secretes mucus and traps air bubbles with its foot, which are stabilized by the amphiphilic mucus. This mucus has evolved from the spawning material of benthic snails, such as wentletraps (*Epitoniidae*), which lay their eggs on an elastic string[21]. Glycosylated proteins in mucus are exploited for various purposes in the animal kingdom, e.g. by snails for locomotion, as discussed above, or as gelling agent for defense by hagfish, as addressed below. The violet sea snail exploits the amphiphilic properties of mucus to stabilize the air-water interface by decreasing the surface tension and forming a viscoelastic interfacial network. The mucus-stabilized foam enables the snail to travel long distances at minimal energy cost.

*Movement by granular rheology.* An exceptional way of locomotion is observed for some snakes and lizards, which show swimming-like movement despite living at the driest places on earth. Granular materials like sand can flow like a fluid, giving rise to this unique way of locomotion (Figure 2c). Animals inhabiting granular media have adopted the ability to fluidize locally the granular material surrounding them. For certain species, digging has proven to be a favorable escape strategy, while others bury their eggs in the sand to protect them from predators and changing climate[25–27]. We focus here on the locomotion of sandfish *(Scincus scincus)*. The mechanical properties of granular media can change dramatically with differing water contents from dry to saturated conditions[28]. A popular way to describe granular flow is frictional rheology[29,30]. In dry conditions, the macroscopic friction is characterized by a dimensionless inertial number, in wet conditions, it is replaced with a dimensionless viscous number[31]. The dimensionless inertial or viscous number is then only a function of the local solid packing fraction.

The sandfish swims in its self-generated (exogen abiotic) localized frictional fluid. Hosoi and Goldman[25] identified four different regimes of locomotion in granular media depending on the dimensionless digger size and inertial number, in which sandfish belong to the large and fast digging organisms. In analogy to low Reynolds number swimmers, the sandfish achieves a net forward displacement by undulating motion. The scallop theorem postulates that at low Reynolds numbers a simple symmetrical back and forth motion is not sufficient for locomotion[32]. However, Qiu *et al.* were able to show that a back-and-forth movement can cause a net forward displacement, provided the swimmer is immersed in a non-Newtonian liquid[33]. The viscosity in the front and rear part of the swimmer varies, thus enabling the forward movement. This means that, in addition to breaking the symmetry by undulating movements, the sandfish also achieves net forward movement by lateral forces that are stronger than the longitudinal forces when moving through granular material (Figure 2c). Movement at a 30 - 45° angle of attack to the longitudinal axis of the sandfish contributes to the forward thrust by Coulomb friction in the surrounding medium. The locomotor's behavior remains the same even at changing packing density[25]. Drag increases linearly with depth, limiting the habitable zone for the sandfish. In contrast, drag increases non-linearly with moisture content due to suction forces, limiting sand-swimming animals to arid regions[34].

**Rheology for attack**
The manipulation of fluid flow for prey is omnipresent in aquatic environments, e.g. for the suction feeding of predatory fish[35] or flow-controlling jellyfish[36] and copepods[37], whereas land-living animals generally rely on endogen biotic complex fluids. Herein we demonstrate how complex fluids can be used to capture prey exploiting viscoelasticity and the flow properties of granular media (Figure 3).

*Preying with complex fluids.* One strategy is the immobilization of prey by strain-hardening and adhesive fluids, as applied by velvet worms (*Onychophora*). The 0.5 - 20 cm long, many-legged velvet worms inhabit humid regions of the tropical and temperate zone[38]. Velvet worms have evolved a predatory lifestyle and as they need to remain moist, and foraging is thus limited to nighttime. The few hours available for foraging favored a carnivorous diet with a low cost/benefit ratio[39]. Further, velvet worms are not particularly fast, and only small prey is catchable without the use of slime. Velvet worms have adapted to this evolutionary pressure by using a endogenous complex fluid for immobilizing prey.

Velvet worms sneak up on their prey and eject a sticky slime from two oral papillae flanking their mouth, as depicted in Figure 3a. The slime ejection induces uncontrolled papillae oscillations at 30 - 60 Hz, which favor the crossing of threads mid-air leading to an entangled slime network immobilizing the prey[40,41]. The velvet worm then injects its hydrolytic, enzyme-containing saliva, killing the prey and inducing liquefaction for later ingestion[39,42]. In contrast to other excreted adhesive fibers in the animal kingdom like spider or silkworm silk, which are solid upon excretion, velvet worm slime is truly a remarkable showcase of a biological complex fluid. Velvet worm slime is fluid at rest, and only develops its cohesiveness and mechanical strength upon elongational spinning[43,44]. The main functional components of the slime, which contains 90% water, are lipid droplets covered by proteins[43]. The oil droplets are approximately 150 nm in diameter with narrow size distribution and are stabilized from aggregation by electrostatic repulsion induced by the charged proteins. We thus propose that the crude slime can be considered a protein-stabilized, oil-in-water nanoemulsion, which is in-line with Baer et al. [45]. The same authors[46] recently demonstrated the effect of varying pH and ionic strength on these oil droplets, which is indeed comparable to protein-stabilized emulsion droplets[47]. Analyses of the adsorbed proteins revealed that they show high proline and charge content and contain collagen-like domains[48–50]. The proteins were long thought to lack any secondary structure, however, more recently Baer *et al.* reported a significant β-sheet content[50].

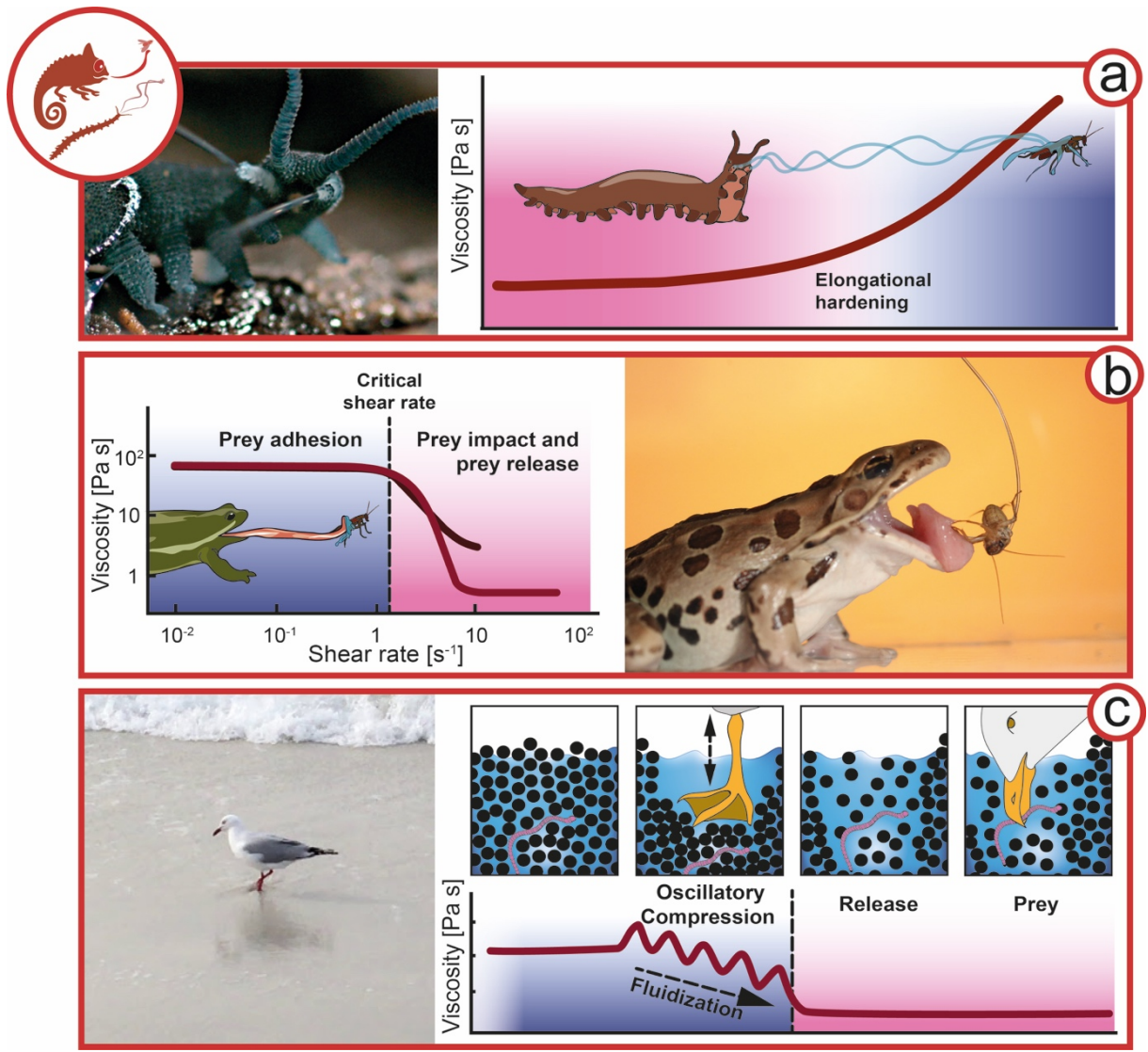

**Figure 3: Exploiting complex fluids for catching prey.** (a) Slime ejection from oral papillae of a velvet worm (*Principapillatus hitoyensis*, Onychophora) from Baer *et al.*[43]. Proposed viscosity change during the elongation ejection of velvet worm slime. (b) Viscosity as a function of the shear rate (re-drawn from Noel *et al.*[51]). Frog (*Anura*) capturing an insect. (c) Seagull (*Laridae*) feeding on worms by fluidizing sand. Proposed rheological behavior of wet sand under oscillatory compression by a seagull preying on worms. Rights and permission: velvet worm (a) adapted with permission from Baer *et al.*[43]. (Copyright 2019 American Chemistry Society) and frog (b) adapted with permission from Noel *et al.*[51]. (Copyright 2019 Royal Society).

The transition of fluid-like crude slime to elastic threads upon extrusion suggests that elongational hardening is crucial for slime functionality, i.e., the viscosity of the slime increases upon elongation (proposed rheology in Figure 3a). The molecular transition during slime elongation was recently elucidated[43]. Upon elongation, the protein-covered oil droplets are elongated into long fibers with a protein core and a thin lipid coating, giving rise to the assumption that the fat droplets are broken up during spinning and the protein fibers are responsible for the slime elasticity[43]. The final slime consists of thin, elastic threads with cohesive droplets distributed along the threads. Upon drying, the threads' tensile strength increases and reaches 4 GPa for the fully dried fibers due to a glass transition[43,48]. Stunningly, the initial protein-covered lipid droplets are reformed upon rehydration of the dried slime, and new fibers can be drawn from regenerated slime[43,44]. It was concluded that intermolecular interactions in velvet worm slime are non-covalent, but rather of hydrophobic or ionic nature.

*Viscous adhesion by frogs and chameleons.* Chameleons (*Chamaeleonidae*) and frogs (*Anura*) use specialized saliva for attack (Figure 3b). Chameleon saliva has a viscosity 400 times higher than human saliva. As a consequence, prey adheres to chameleon tongues by viscous adhesion[52]. For this attack strategy, the viscous adhesion to the tongue needs to exceed prey inertia. However, due to the high viscosity of chameleon saliva, the prey size is not limited by viscous adhesion[53]. Similarly, frogs possess a viscoelastic tongue coupled with high viscosity and shear-thinning saliva (Figure 3b). Upon tongue impact, a critical shear rate is exceeded, which decreases the viscosity and ensures the spreading of saliva. The viscoelastic tongue acts as a shock-absorber and adapts to prey shape, further increasing the tongue-prey contact area. A lower shear rate is applied by tongue retraction, resulting in a higher saliva viscosity[51]. Hence, the saliva must recover sufficiently fast to provide viscous adhesion.

*Exploiting granular rheology for prey.* A specialized technique exploiting granular rheology can be observed for seagulls (*Laridae*) in tidal zones in the form of two-footed pedaling. Various glovers and gulls pedal the sand to loosen its structure around worms and ease their prey[54]. It was shown that on solid dry ground, the vibrations mimic approaching moles and promote the escape of worms[55]. On wet ground, the two-footed pedaling has been shown to promote the release of small animals by the fluidization of the wet sand[56,57]. Here we propose a mechanism based on granular rheology (see Figure 3c). Defined rheologically, the sand in tidal zones is randomly close-packed and completely wetted with a glossy surface. Upon pedaling, the sand is spatially rearranged and the water table is lowered temporarily; the sand surface appears matte. When the surrounding water refills the void, the sand is diluted and can flow with far less resistance due to its lower solid volume fraction.

The seagull is then able to pick the prey from the diluted suspension rather than from densely packed sand with high resistance to deformation. Hence, the pedaling, representing oscillatory compression, results in a more dilute sand structure (positive dilatancy[58]).

**Rheology for reproduction & parental care**

Complex fluids play a crucial role in animal reproduction. While it is well-known that the rheology of cervical mucus is vital for sexual reproduction in humans[59], many animals use external complex fluids for reproduction and parental care (Figure 4).

*Viscoelasticity in courtship.* Probably the most spectacular example of complex fluids for reproduction is the eerie, but beautiful, mating ritual of leopard slugs (*Limax maximus*). The twosome of leopard slugs use a mucus thread to suspend themselves mid-air to perform their circus-like sexual act[60] (Figure 4a). The mucus is usually used for adhesive locomotion by the slugs (as addressed above). Once a male- and female-behaving individual encounter, the female slug will lead the way up a tree. When the desired location is reached, preferably the bottom side of a tree branch, the snails intertwine and form a thick mucus layer around themselves. The slugs rub their secreted mucus against each other for up to one hour[61] until a viscoelastic mucus quality is achieved. The slugs then lower themselves from the branch, dangling mid-air by this mucus.

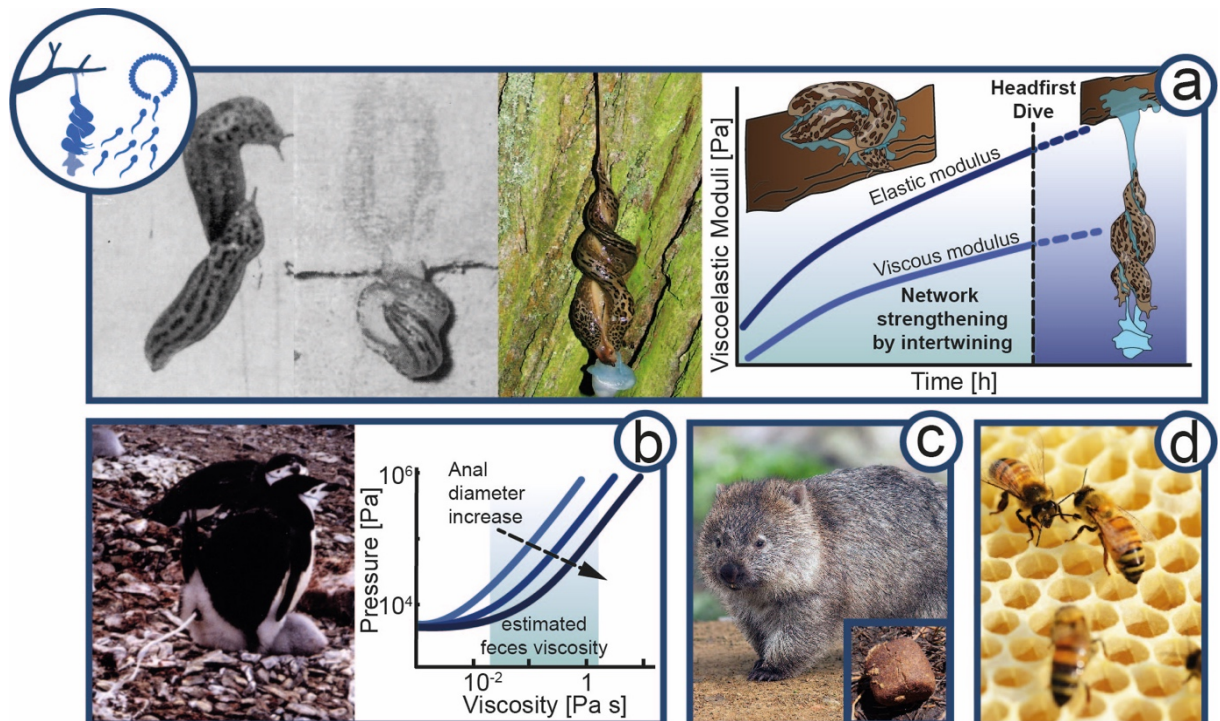

*Figure 4:* **Flow of complex material in reproduction and parental care.** (a) Viscoelasticity in courtship of the leopard slug (*Limax maximus*). Slugs meet mating partner, intertwining of mucus, and hanging intercourse after headfirst dive. Proposed viscoelasticity of *Limax maximus* mucin over time during courtship and mating. (b) Feces rheology of penguin (*Pygoscelis*) housekeeping. Graph is re-drawn from Meyer-Rochow et al., [62]. (c) Wombat (*Vombatidae*) cubic feces to mark territory (d) Bees (*Apis*) constructing hexagonal shaped comb in a hive. Rights and permission: (a)[61] (credit to The Ohio State University and The Ohio Academy of Science) and T.Hiddessen (CC BY-SA 3.0), (b) kindly provided and with permission from V.B. Meyer-Rochow, (c) Wombat and wombat feces: Bjørn Christian Tørrissen (CC BY-SA 3.0) and Sheba Also (CC BY-SA 2.0).

We suggest that time-dependent rheological properties and constant intertwining (shearing) are crucial to this mating ritual. As generally observed for salivary, nose, and snail mucus, the yield stress and viscosity increase upon drying. Mucus viscosity was shown to exhibit a second or third order increase with mucin concentration[63]. The shear stresses induced by the constant intertwining may further promote mucus elasticity by the formation of elastic threads. Shear forces are known to cause aggregation of mucin molecules into fiber-like structures[64]. Thus, the slugs must react to the changing material properties by drying and intertwining to achieve ideal viscoelasticity before performing a slow-motion head dive together. The snails have to time their suspension with perfection. Too early and the snail thread will not hold the weight of two slugs. Wait too long and the mucus will dry out and lose its elastic properties, becoming solid. In both cases, the slugs would not be able to exploit the viscoelasticity to lower themselves from the tree branch.

    A mid-air position allows full extension of male genitalia, a feat difficult to perform without being suspended. The suspending mucus thread can be up to 50 cm in length[61]. The mucus thread will experience a strong extensional force as it holds two fully-grown leopard slugs, estimated to weigh around 1 to 8 g each[65]. This gravitational force has to be balanced by the elasticity of the mucus thread. At the end of the sexual encounter, the thread may either rupture before the snails leave or, in some cases, the snails eat up their mucus[61].

*High pressure extrusion for defecation.* Penguins (*Pygoscelis*) often lay eggs at low temperatures that do not permit leaving the egg unprotected even for the blink of an eye. Penguins incubate in shifts of several days allowing the other parent to feed[66]. To avoid exposure of the egg or newborn and prevent contamination, chinstrap *(P. antarctica)* and Adélie (*P. adeliae*) penguins propel their feces radially from the nests (Figure 4b), which are even at high population density no closer than 60 cm[67,68]. With an approximate firing distance of 40 cm, defecation cannot reach neighboring nests and thus prevent soiling of its inhabitants. Meyer-Rochow and Gal[62] estimate the expulsion pressure at the *orificium venti* from the density and viscosity of penguin feces, firing range, shape, aperture, and height of the anus. Depending on food sources, color and viscosity may vary from white to pinkish and from a few millipascal (watery consistence) up to 110 mPa (olive oil), respectively. The velocity of the feces is at 2 m/s with an average volume of 20 mL and a firing time of 0.4 s. Summing up the pressure required to accelerate the fluid volume and the pressure needed to overcome the viscous friction (Hagen-Poiseuille equation) leads to an expulsion pressure of up to 10 to 60 kPa or about 1000 mmHg, which is about four or eight times the blood pressure of giraffes or humans, respectively. The pressure curves are shown for three cloacal apertures (Rockhopper (*Eudyptes chrysocome*) = 4.2 mm, Adélie = 8.0 mm, and Gentoo (*Pygoscelis papua*) = 13.8 mm). Calculations are impaired by viscosity measurements of the feces due to low sample volume and a rather inhomogeneous sample constitution (e.g., compromised by remnants of crustacean cuticles and other particles). Further, at the given viscosity, density, and firing velocity, turbulent flow might occur at feces viscosities higher than 90 mPa, which would require substantially higher pressures to reach the same distance than under laminar flow conditions. In summary, the feces viscosity and the cloacal apertures allow the penguin to defecate without leaving the nest, allowing the penguin to efficiently hatch and protect their offspring.

*Cubic feces as communication.* Another curious fecal secretion is observed for wombats (*Vombatidae*) (Figure 4c). Of all animals, wombats are the only ones that secrete excrement in a cubic form[69,70]. It is suggested, that wombats use their feces as a communication tool to mark their territory and to communicate with each other by scent. The cubic shape of wombat feces could prevent it from rolling off the stones and cliffs, habitats where wombats often live. The feces are produced through alternating stresses along the gastrointestinal tract. As the interior intestinal wall is less elastic than the exterior, the exterior wall is stretched more and feces are formed in squares.

*Foam for spawn protection and aeration.* Fish and amphibia reproduce by spawning, i.e., the release or deposition of eggs in an aquatic environment. A subfamily of armored catfish (*Callichthyinae*) and a family of frogs (*Rhacophoridae*) further protect their spawn by the formation of protective foam. Foam formation requires a surface-active substance that stabilizes the energetically unfavorable air-water interface by a decrease in surface tension[71]. The catfish do so by swimming in circles belly-up close to the water surface and pumping water through the gills[72]. In the gills, the water is enriched with amphiphilic mucin that acts as surface-active agent and stabilizes the foam. The same mechanism of foam formation is exploited by the violet sea snail for passive flotation, as previously discussed. *Callichthyinae* are thought to have evolved this strategy to improve the aeration of their spawn as they often inhabit oxygen-deprived waters in the tropics[73]. *Rhacophoridae,* on the other hand use a protein, ranaspumin, as stabilizer for their protective foam. The female secrets the amphiphilic ranaspumins which are whipped into a stable foam using their legs, with a method similar to whipping cream. The ranaspumins allow the production of a foam that is stable enough to withstand environmental conditions for at least 10 days. The foam protects the eggs from predation and climate variations while providing sufficient oxygen[74].

*Flow in the hive.* Honeybees (*Apis*) live communally and care for their offspring cooperatively. In their nest, honeybees construct an elaborate system of highly structured, hexagonal, prismatic cells known as honeycomb (Figure 4d). Inside the cells, the larvae are raised and honey and pollen are stored. The formation mechanism and the precise repetitive geometry of the honeycombs intrigued natural philosophers and scientists for centuries and continues unabated[75–79].

    The bees construct the comb cells by secreting endogen biotic thermoplastic wax. Beeswax is a viscoelastic building material that contains more than 300 different chemical components[80]. As for other fatty systems, beeswax does not have a sharp melting point. With increasing temperature wax liquefies, i.e., viscous properties increase and the elasticity decreases[81]. The transition from fully elastic crystalline wax to the viscoelastic amorphous structure is not continuous and takes place in two steps at about 25 and 40°C. Bees can raise their body temperature to more than 43°C, allowing them to change the malleability of the wax[80]. In the early stage of comb construction, bees use their bodies as templates to build cylindrical cells around themselves[77,82,83]. This shape remains stable for many weeks after their construction. The final hexagonal shape is formed when the bees heat the wax to 37 - 40°C and it becomes amorphous and viscoelastic[77]. The decreased viscosity enables the wax to flow into its energetic most favorable shape due to surface tension[82]. The flow into the triple junction between adjacent walls results in straight walls and 120° angles relative to another; the well-known hexagons. The same

process is also observed when foam bubbles get into contact[77,80,82]. Hence, bees shape the hexagonal cell by controlling the viscoelasticity of the endogenous wax over temperature.

**Rheology for defense and protection**
Animals have found a variety of defensive protection strategies exploiting complex fluids, e.g., viscoelastic slime or foam for protection or granular rheology (Figure 5).

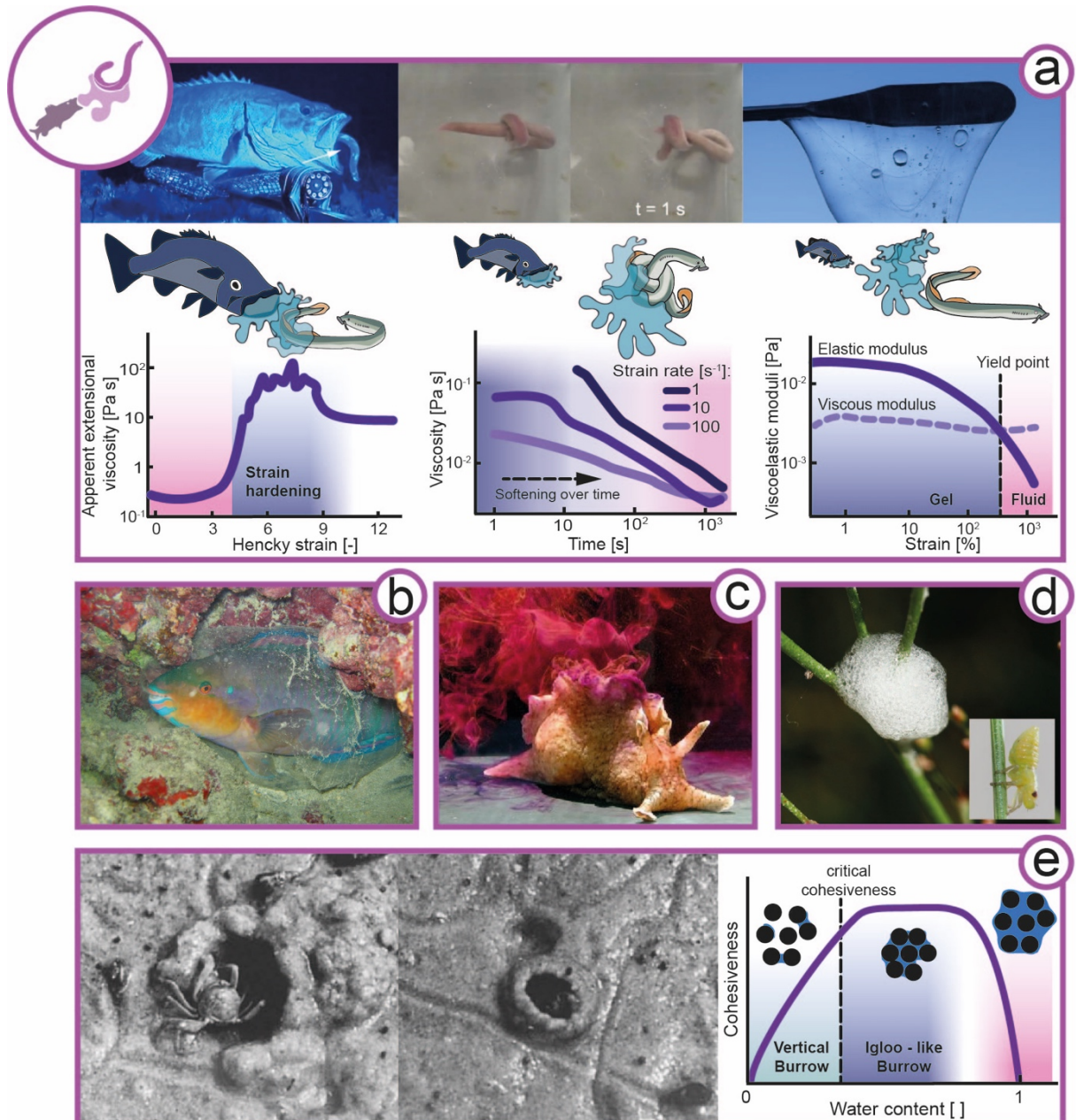

**Figure 5: Rheology of complex flow in defense.** (a) Hagfish (*Myxine glutinosa, Eptatretus stoutii*) slime formation for defense and knot formation to escape its slime. Apparent extensional viscoelasticity of hagfish mucin (left), viscosity of hagfish slime under shear (middle), and viscoelasticity of hagfish slime (right) (redrawn from Böni *et al.,*[64]) (b) Mucus sleeping bag of the parrot fish (*Chlorurus sordidus*). (c) Inking sea hare as a defensive slime (*Aplysia californica*). (d) Protein-stabilized foam for protection from predators by spittlebug (*Cercopidae*) nymphs. (e) Sand crab (*Dotilla*) building an igloo or a hole depending on the water content in the sand. Suggested cohesiveness of sand with increasing water content. Rights and permissions: (a) From Zintzen *et al.*[84] and re-drawn from Böni *et al.*[64] (CC BY-NC-ND 3.0). (b) From Igor Cristino Silva Cruz (CC BY-SA 4.0), (c) With kind permission from Genevieve Anderson, (d) David Iliff (CC BY-SA 3.0) and Dave (CC BY-SA 3.0), (e) adapted with permission from Takeda et al.[85] (Copyright 2019 Elsevier).

*Viscoelastic gel to avoid predation.* One particularly striking example of animal defense using complex fluids is the notorious hagfish (*Myxine glutinosa, Eptatretus stoutii*) (Figure 5a). Hagfish are crucial for aquatic ecosystems as their burrowing and feeding activities have a significant impact on substrate turnover and ocean cleanup by feeding on carcasses that sink to the sea bed[86]. When hagfish are attacked by predators such as sharks or suction feeding fish, they form vast amounts of slime in less than a second. This remarkable defensive slime formation is triggered by the release of hagfish exudate from specialized pores into the surrounding seawater. Once in contact with water, the exudate rapidly forms a fibrous hydrogel that clogs the mouth and gills of the predator[87,88]. The crude exudate is composed of coiled-up skeins and mucin vesicles, both crucial components for slime formation. Hagfish skeins are keratin-like protein threads coiled up into a microscopic ball of yarn with dimensions of 50 by 150 $\mu$m.[89,90] When hagfish skeins come in contact with water, the skeins unravel into long protein threads with a length of up to 15 cm[87,91,92]. The mucin vesicles swell in contact with water and eventually burst, releasing water-absorbing mucin molecules into the slime network[93]. As opposed to conventional hydrogel formers, mucins gel without additional energy input and swell immediately.

Hagfish slime forms a very soft, yet elastic hydrogel with a higher water content than any other known biological hydrogel[64,94]. A functional slime can only be formed when both components are present in their natural environment. The long skein threads are crucial for cohesion and viscoelasticity of the slime, whereas the mucins facilitate water entrapment[64,87]. The hagfish slime is unique because it gels large amounts of cold water within seconds. As only very little exudate (approximately 0.02 wt% solids in final slime), and no further energy input is required, the slime is an economical yet efficient defense mechanism. Due to the low solid content the slime is short-lived, mechanically unstable, and prone to wash out. However, as rapid gelling is more important than longevity to avoid seizure, the slime seems perfectly timed to the defense mechanism of the hagfish[64,84].

The rheology of hagfish slime is fine-tuned for the desired functionality. Hagfish mucins alone exhibit time dependent viscosity; the viscosity decreases as a function of time (thixotropy) in agreement with the short-lived nature of the slime.[64] In extensional flow, mimicking the suction flow of suction feeding fish, the mucin viscosity increases. Hagfish slime causes the viscosity to shoot up by two orders of magnitude in just a second, efficiently deterring suction feeding predators[64]. On the other hand, the slime is thinning under shear, and the hagfish can easily wipe slime off itself by the formation of a knot that is sheared down its body[95]. This mechanism allows the hagfish to escape its trap after successfully deterring the predator.

*Viscoelastic sleeping bags, distracting taste, and slimy stars.* The hagfish is not the only sea creature that uses slime to deter its predators. The slime star (*Pteraster tesselatus*), uses respiratory water flow and production of mucus to produce large quantities of slime when molested[96,97]. Water is pressed through mucus channels, rapidly forming a slime body, engulfing the sea star. There is currently no rheological data available on this slime, but we suggest that the viscosity and viscoelasticity could be similar to the hagfish.

The parrotfish (*Chlorurus sordidus*)[98] secretes mucus at night to form a gelatinous, protective sleeping bag around its body to protect itself from ectoparasitic gnathiid isopods (Figure 5b). Other protective secretions are made by sea hares (*Aplysia californica*). Sea hares are not on the dietary plan of most marine animals, due to two unpalatable secretions, ink and opaline, which sea hares squirt at approaching predators (Figure 5c). These viscous secretions stick to the antennules and mouth of predators. Opaline further sticks to the chemosensors and physically blocks the perception of food odors. The high viscosity mimics the feedings stimulus and the concentrated amino acids in the secretion induce an overstimulation of the chemosensors. These combined mechanisms result in attacker retreat[99,100].

*Foam for protection.* The spittlebug (*Cercopidae*) nymphs produce foam for protection from predators, moisture loss, UV-radiation, and temperature variations, as shown in Figure 5d[101–103]. As the ancestor of the spittlebug lived underground, it is believed that the foam could have enabled the spittlebug to adopt a lifestyle above ground[101]. The foam production of spittlebug nymphs is further linked to their feeding on xylem sap. The sap transported in plant xylem is much lower in nutrients than phloem sap, resulting in excess liquid uptake[104]. The foam is stabilized by a secreted protein[38], similar to the foam produced by *Rhacophoridae* discussed above. This secreted protein stabilizes the water-air interface through a predominantly viscoelastic network.

*Sand igloos.* Sand-dwelling crabs of the genus *Dotilla* have been observed to create vertical burrows or igloo-like structures depending on the water content of sand (Figure 5e)[85,105,106]. The water content significantly alters the cohesiveness of sand by a shift from dominating friction forces to arising suction forces. When sand is firm and well-drained, the crabs create vertical burrows. However, under unstable and semi-fluid conditions, they adapt their behavior and create igloo-like structures. As demonstrated by Takeda et al.,[85] the sand has to be semi-fluid for the igloo to hold its shape. Under these semi-fluid conditions, the sand would not allow for the construction of a vertical burrow. Hence, crabs exploit the water-induced suction forces within sand particles that provide cohesiveness and allow the construction of self-standing architectures. Note the contrast to sandfish that exploit fluid-like behavior of dry sand for swimming-like locomotion, as discussed above.

**Discussion and conclusion**
In this work, we highlight striking examples of animals using complex fluids as part of their survival strategy. In particular, we demonstrate how certain animals manipulate their surrounding complex fluids or have evolved endogen biotic materials. In this conclusion, we discuss the relevant time scales of complex fluids for animals, the evolution of the most common involved polymer mucin, and the ability of animals to sense environmental rheological conditions.

*Relevant conditions and time scales.* A specialty of complex fluids in the animal kingdom is that animals often have little influence on ambient conditions. Yet in many cases, the fluids need to change their properties significantly within a few seconds. An extreme example is hagfish slime, which manages to gel instantly vast amounts of water despite low water temperatures. Another way to alter fluid properties at given conditions is elongational hardening, as exploited by the velvet worm, transforming their liquid crude slime into elastic fibers by elongation. On a longer time scale of minutes to hours, the leopard slugs let their mucus dry and favor fibrillation by shear. Much longer time-scales are desired for foam structures as introduced for passive flotation or protection. In these cases, amphiphilic mucin or proteins are employed to provide foam stability up to several days.

*Mucin: a molecule with universal applications for animals.* Mucins are glycoproteins, i.e. proteins with covalently bound sugar residues. Mucus, the aqueous mucin-containing solution, has popped up throughout all sections of this review. Besides the special uses of mucus addressed here, it is present in all mucosa and involved in many vital functions of animals like nutrient uptake in the gastro-intestinal system or oxygen transport in lungs. This broad range of applications and versatility of mucus probably derives from its early appearance in evolution. Mucus evolved first in the *Cnidaria phylum* (polyp and medusa), which includes corals and the *Ctenophora phylum* (jelly fish) for ciliary-mucus driven particle feeding and uptake of nitrogen[107]. In the course of time, mucus has evolved into various specialized complex fluids. Mucus can fulfill these versatile tasks due to the broad toolbox of proteins and sugars that can be combined into mucins. Depending on the structure and composition of mucins they can: provide suitable rheology for adhesive locomotion or the adhesive force of a chameleon tongue; act as amphiphilic substances stabilizing foams for passive flotation or protection; facilitate incorporation of water in hagfish slime; or facilitate the spectacular mating ritual of leopard slugs.

*The sixth sense.* In various situations animals adapt their behavior to given material properties. For example, sand crabs adapt their burrow shape to the water content of the sand, thereby reacting to changes in material properties due to a shift between suction and frictional forces. A skill that human offspring often lack in their attempts at building complex sand structures. The highest level of trust in their rheology skills is certainly required by the leopard slug twosome, which needs to time its headfirst dive perfectly to complete their mating ritual without falling.

To conclude, various animals have adopted survival strategies that exploit complex flow of endogen or exogen fluids to gain an evolutionary advantage (Darwinian fitness). Further, there are several indications that animals can adapt their behavior to varying material properties, revealing a sixth sense for complex fluid rheology. Hence, rheology or materials science in general, help to understand evolution and animal behavior and provide a quantitative approach towards ethology.

**Acknowledgments**
The authors acknowledge the Swiss National Science Foundation for funding, project Nos. P300P2_171233 and 200021-175994, and Caroline Giacomin for proof-reading.

Animal Rheology

Animal Rheology64. Böni, L., Fischer, P., Böcker, L., Kuster, S. & Rühs, P. A. Hagfish slime and mucin flow properties and their implications for defense. *Sci. Rep.* **6**, 30371 (2016).
65. Barker, G. M. & McGhie, R. A. The biology of introduced slugs (Pulmonata) in New Zealand 1. Introduction and notes on Limax maximus. *New Zeal. Entomol.* **8**, 106–111 (1984).
66. Numata, M., Davis, L. S. & Renner, M. Prolonged foraging trips and egg desertion in little penguins (Eudyptula minor). *New Zeal. J. Zool.* **27**, 277–289 (2000).
67. Wiliams, T. D. *The penguins. Spheniscidae*. (1991).
68. Rietkerk, M. & van de Koppel, J. Regular pattern formation in real ecosystems. *Trends Ecol. Evol.* **23**, 169–175 (2008).
69. Patricia, J Yang, Miles Chan, Scott Carver, D. L. H. How do wombats make cubed poo? in *http://meetings.aps.org/Meeting/DFD18/Session/E19.1* (2018).
70. Allen, M. Why wombats have cube-shaped poo. *Phys. World* **32**, 5 (2019).
71. Walstra, P. Principles of Foam Formation and Stability - Foams: Physics, Chemistry and Structure. in (ed. Wilson, A.) (Springer London, 1989).
72. Andrade, D. V & Abe, A. S. Foam nest production in the armoured catfish. *J. Fish Biol.* **50**, 665–667 (1997).
73. Mol, J. H. A. Structure and function of floating bubble nests of three armoured catfishes (Callichthyidae) in relation to the aquatic environment - The Freshwater Ecosystems of Suriname. in (ed. Ouboter, P. E.) (Springer Netherlands, 1993).
74. Cooper, A. *et al.* Adsorption of Frog Foam Nest Proteins at the Air-Water Interface. *Biophys. J.* **88**, 2114–2125 (2005).
75. Hunter, J. VIII. Observations on bees. *Philos. Trans. R. Soc. London* **82**, 128–195 (1792).
76. Waterhouse, G. R. II.: On the Formation of the Cells of Bees and Wasps. *Trans. R. Entomol. Soc. London* **12**, 115–129 (1864).
77. Pirk, C. W. W., Hepburn, H. R., Radloff, S. E. & Tautz, J. Honeybee combs: construction through a liquid equilibrium process? *Naturwissenschaften* **91**, 350–353 (2004).
78. Narumi, T., Uemichi, K., Honda, H. & Osaki, K. Self-organization at the first stage of honeycomb construction: Analysis of an attachment-excavation model. *PLoS One* **13**, e0205353 (2018).
79. Zhang, K., Duan, H., Karihaloo, B. L. & Wang, J. Hierarchical, multilayered cell walls reinforced by recycled silk cocoons enhance the structural integrity of honeybee combs. *PNAS* **107**, 9502–9506 (2010).
80. Tautz, J. *The Buzz about Bees*. (Springer Berlin Heidelberg, 2008).
81. Shellhammer, T. H., Rumsey, T. R. & Krochta, J. M. Viscoelastic properties of edible lipids. *J. Food Eng.* **33**, 305–320 (1997).
82. Karihaloo, B. L., Zhang, K. & Wang, J. Honeybee combs: how the circular cells transform into rounded hexagons. *J. R. Soc. Interface* **10**, (2013) 20130299.
83. Nazzi, F. The hexagonal shape of the honeycomb cells depends on the construction behavior of bees. *Sci. Rep.* **6**, 28341 (2016).
84. Zintzen, V. *et al.* Hagfish predatory behaviour and slime defence mechanism. *Sci. Rep.* **1**, 131 (2011).
85. Takeda, S., Matsumasa, M., Yong, H.-S. & Murai, M. "Igloo" construction by the ocypodid crab, Dotilla myctiroides (Milne-Edwards) (Crustacea; Brachyura): the role of an air chamber when burrowing in a saturated sandy
19